\input{aipcheck}

\documentclass[
    ,final            
  ]
  {aipproc}

\layoutstyle{6x9}

\begin{document}

\title{Soft Collinear Effective Theory}


\author{Sean Fleming}{
  address={Physics Department, University of Arizona, Tucson, AZ 85719}
}

\begin{abstract}
In this talk I review soft collinear effective theory. After a discussion of the formalism and properties of the effective field theory, I turn to phenomenology. I present results on color-suppressed $B \to D$ decays, and on the $\Upsilon$ radiative decay spectrum.
\end{abstract}

\maketitle

\section{Introduction}

Effective field theories provide a simple and elegant method
for calculating processes with several relevant energy
scales
\cite{Weinberg:1978kz,Witten:kx,Georgi:qn,Harvey:ya,Manohar:1995xr,Kaplan:1995uv}.  
Part of the utility of effective theories is that 
they dramatically simplify the summation of logarithms of
ratios of mass scales, which would otherwise make perturbation theory
poorly behaved.  Furthermore the systematic power counting in effective
theories, and the approximate symmetries of the effective field theory
can greatly reduce the complexity of calculations.

In this talk I review soft-collinear effective theory (SCET)~\cite{Bauer:2000ew,Bauer:2000yr,Bauer:2001ct,Bauer:2001yt}, which is an effective field theory describing the dynamics of highly energetic particles moving close to the light-cone interacting with a background field of soft quanta. 

\section{Soft collinear effective theory}

\begin{figure}\label{bm}
\includegraphics[height=.3\textheight]{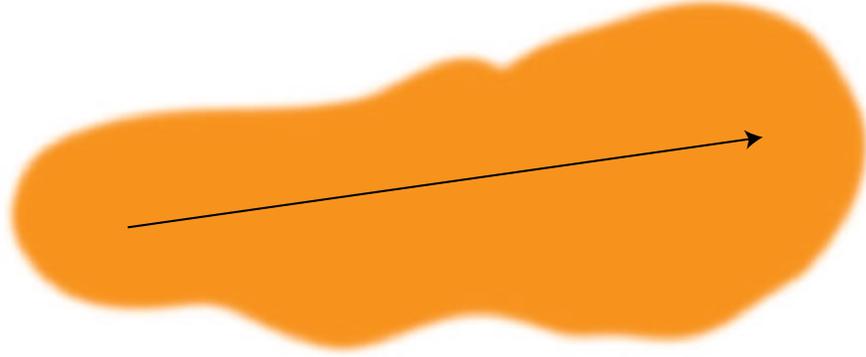}
\caption{Collinear particle traveling through brown muck.}
\end{figure}

\begin{figure}\label{bm2}
\includegraphics[height=.25\textheight]{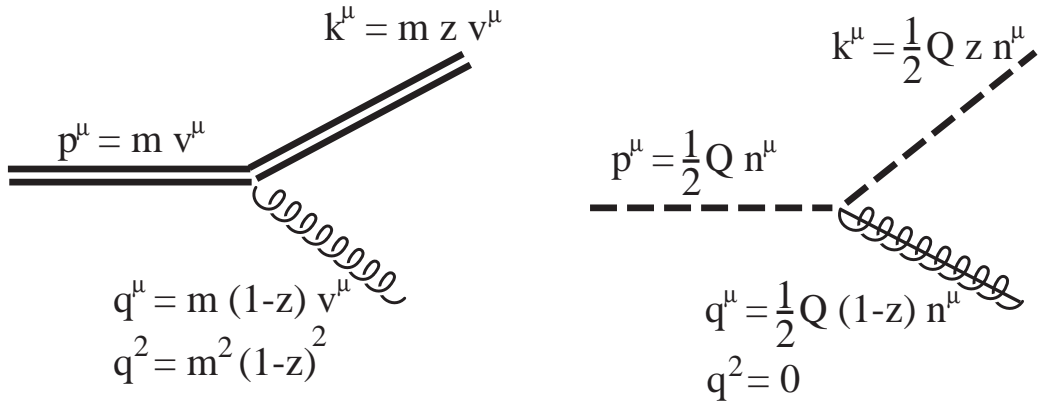}
\caption{HQET splitting not allowed for arbitrary $z$ (left). SCET collinear splitting allowed for any $z$ (right).}
\end{figure}

A simple picture of the types of process which SCET applies to is given in Fig.~\ref{bm}. Here the brown muck represents a background of soft particles  ({\it i.e.} particles with momenta of order $\Lambda_{\textrm{{\small QCD}}}$) through which travel collinear degrees of freedom with  large energy $Q \gg \Lambda_{\textrm{{\small QCD}}}$. 
The interaction of the collinear particle with the background introduces a small residual momentum component into the light-like collinear momentum so that collinear particles in the figure have momentum $p^\mu = Q n^\mu+k^\mu$,  where $k^\mu \sim \Lambda_{\textrm{{\small QCD}}}$ and $n^\mu = (1,0,0,-1)$. This can be compared with heavy quark effective theory (HQET) where the $b$ quark inside the B meson has momentum $p^\mu_b = M v^\mu + k^\mu$ where $v^\mu = (1,0,0,0)$. There is however, a difference between heavy particles in HQET and collinear particles in SCET. The class of interactions shown on the left-hand side Fig.~\ref{bm2}  are not allowed in HQET for arbitrary values of the momentum fraction $z$, since that would require the presence of a gluon with typical invariant mass of order $M$, and these have been integrated out. In SCET these types of interactions, shown on the right-hand side of Fig.~\ref{bm2},  are allowed since they do not require the gluon to have a large invariant mass: any collinear particle can decay into any number of collinear particles. As a consequence the SCET Lagrangian is more complicated than the HQET Lagrangian.
In particular SCET consists of two sectors: soft and collinear. I give the leading SCET Lagrangian for the quarks  without derivation, since this has been covered extensively in the literature, first in a label  formulation~\cite{Bauer:2000ew,Bauer:2000yr,Bauer:2001ct,Bauer:2001yt}, and subsequently in a position space formulation~\cite{Beneke:1,Beneke:2}. The part of the SCET Lagrangian involving collinear interactions can be split into two pieces: one which includes the coupling of collinear quarks to soft gluons
\begin{equation}\label{scetlag}
{\cal  L}_{cs} = \bar{\xi}_{n,p'}  i n\cdot D  \frac{{\bar n\!\!\!\!\slash}}{2} \xi_{n,p} \,,
\end{equation}
where $i n\cdot D = i n \cdot \partial +  g n\cdot A_{s}$, with $A_s$ the soft gluon field. This expression looks very much like the HQET Lagrangian with the velocity $v^\mu$ replaced with the light-like vector $n^\mu$, and was first derived in Ref.~\cite{Dugan:1990de}. The second piece of the collinear Lagrangian consists of interactions of only collinear particles among themselves:
\begin{equation}
{\cal  L}_{c} = \bar{\xi}_{n,p'} \bigg\{  g n\cdot A_{n q} + i {D\!\!\!\!\!\!\slash}^\perp_c
\frac{1}{i\bar{n} \cdot D_c} i   {D\!\!\!\!\!\!\slash}^\perp_c \bigg\} \frac{{\bar n\!\!\!\!\slash}}{2}
\xi_{n,p}
\,.
\end{equation}
There are some very interesting symmetries exhibited by SCET. Though it is not obvious from the expression above, the Lagrangian is invariant under separate collinear and soft gauge transformations which provides a powerful restriction on the operators allowed in the theory~\cite{Bauer:2001yt}. Furthermore the Lagrangian is invariant under a global $U(1)$ helicity spin symmetry, and must be invariant under certain types of reparameterizations of the collinear sector of the Lagrangian~\cite{Manohar:2002fd,Chay:2002mw}. 

\section{Color-Suppressed $B \to D$ Decays}

Next I turn to phenomenology and review an SCET calculation of color-suppressed $B \to D$ decays, see Ref.~\cite{Mantry:2003uz}, and related papers~\cite{Mantry:2004pg,Blechman:2004vc,Leibovich:2003tw}. The reason these decays are so interesting is that in SCET the leading contributions are power suppressed, which makes their structure more complicated than leading order processes. 

In general  $B \to D$ decays can be classified into three three topologies commonly referred to as tree (T), color-suppressed (C), and exchange (E). Both C and E type decays are suppressed by $1/N_c$ in large $N_c$ counting. Decays which fall into these categories are $B^- \to D^0 \pi^-$ (C type), $\bar{B}^0 \to D^+ \pi^-$ (E type), and $\bar{B}^0 \to D^0 \pi^0$ (both C and E type). Of the color-suppressed decays only $\bar{B}^0 \to D^0 \pi^0$ has been observed so far. The  branching ratio is $(0.29 \pm 0.03) \times 10^{-3}$, which is nearly a factor of 50 smaller than T type decay branching ratios. Thus color-suppressed decays are indeed suppressed, but by an amount that can not be explained by a factor of $1/N_c$. SCET can explain the additional suppression: both C and E type decays receive their first contributions at subleading order in the power counting leading to an additional suppression factor of $\sim 0.2$. 

Using the properties of SCET the amplitude for color-suppressed decays can be factored:
\begin{eqnarray}
A^{D^{(*)}}_{00} &=& N^{(*)}_0 \int d x dz dk^+_1 dk^+_2 T^{(i)}(z) J^{(i)}(z,x,k_1^+,k_2^+) s^{(i)}(k_1^+, k_2^+) \phi_M(x) 
\nonumber \\
& & + A^{D^{(*)}M}_{\textrm{\tiny long}} \,.
\end{eqnarray}
The hard kernel $T^{(i)}(z)$ includes physics at the scale $Q^2 \sim m^2_b$, and the jet function $J^{(i)}(z,x,k_1^+,k_2^+)$ includes physics at the scale $Q\Lambda$, where $\Lambda \sim 1 \, \textrm{GeV}$. Both of  these functions are perturbatively calculable. The functions $s^{(i)}(k_1^+, k_2^+)$ and $\phi_M(x)$ are non-perturbative. The latter is the familiar meson light-cone wave function, and the former is the lightcone distribution function of the spectator quarks in the $B$ and $D$ mesons. It is an unknown nonperturbative function which depends on the velocities of the heavy quarks $v,v'$ and the collinear direction  $n$ of the light meson. This implies that the function is universal to both $D$ and $D^*$, and as a result the decay rate for $\bar{B}^0 \to D^0 \pi^0$ and $\bar{B}^0 \to D^{*0} \pi^0$ are the same  up to higher order corrections. The experimental data on the branching fractions is~\cite{Coan:2001ei,Abe:2001zi}
\begin{eqnarray}
Br(D^0 \pi^0 ) & = & (0.29 \pm 0.03) \times 10^{-3}
\nonumber \\
Br(D^{*0} \pi^0 ) & = & (0.26 \pm 0.05) \times 10^{-3} \,.
\end{eqnarray}
Furthermore the function $s^{(i)}(k_1^+, k_2^+)$ is complex and as a result contains a naturally large strong phase. The measurement of the strong phase gives
\begin{eqnarray}
\delta(D^0 \pi^0 ) & = & 30.4 \pm 4.8^\circ
\nonumber \\
\delta(D^{*0} \pi^0 ) & = & 31.0 \pm 5.0^\circ \,.
\end{eqnarray}
The data seems to be a resounding confirmation of the SCET results. There are more predictions made in Ref.~\cite{Mantry:2003uz} and it will be interesting to see if these are confirmed.

\section{$\Upsilon$ Radiative Decay}

Next I turn to the radiative decay of the $\Upsilon$. In a series of papers~\cite{Bauer:2001rh,Fleming:2002rv,Fleming:2002sr,Fleming:2004rk,Fleming:2004hc} NRQCD was combined with SCET to determine the inclusive decay spectrum as well as the exclusive decay rate to $f_2$.

The radiative decay $\Upsilon\to X\gamma$ was first investigated about
a quarter century ago \cite{firstRad}.  The conventional wisdom
was that this process is computable in perturbative QCD due
to the large mass of the $b$ quarks.  Since then, we have learned much
about quarkonium in gerneral \cite{bbl} and this process in particular
\cite{Catani:1995iz,Maltoni:1999nh,Rothstein:1997ac,
Kramer:1999bf,Bauer:2001rh}.  In addition, CLEO will soon update their original measurement of this decay \cite{Nemati:1996xy}.
It is thus timely to reexamine the theoretical predictions for this
rate.

Before the work of Refs.~\cite{Bauer:2001rh,Fleming:2002rv,Fleming:2002sr,Fleming:2004rk,Fleming:2004hc} the standard method for calculating the direct radiative decay of the
$\Upsilon$ was the operator product expansion (OPE) with 
operators scaling as some power of the relative velocity of the heavy
quarks, $v$, given by the power counting of Non-Relativistic QCD
(NRQCD) \cite{bbl}.  The $v\to0$ limit of NRQCD coincides with the
color-singlet (CS) model calculation of~\cite{firstRad}.
This picture is only valid in the intermediate range of the photon
energy ($0.3 \leq z \leq 0.7$, where $z = 2 E_\gamma
/M$, and $M = 2m_b$).  In the lower range, $z\leq 0.3$,
photon-fragmentation contributions are important~\cite{Catani:1995iz,
Maltoni:1999nh}. At large values of the photon energy, $ z\geq
0.7$, both the perturbative expansion~\cite{Maltoni:1999nh} and the
OPE~\cite{Rothstein:1997ac} break down.
This breakdown is due to NRQCD not including collinear
degrees of freedom.  The correct effective field theory is a
combination of NRQCD for the heavy degrees of freedom and SCET for the light
degrees of freedom.

In Refs.~\cite{Bauer:2001rh,Fleming:2002rv,Fleming:2002sr,Fleming:2004rk}
the different contributions were combined to obtain a prediction for
the photon spectrum, which is shown
\begin{figure}
\includegraphics[height=.7\textheight]{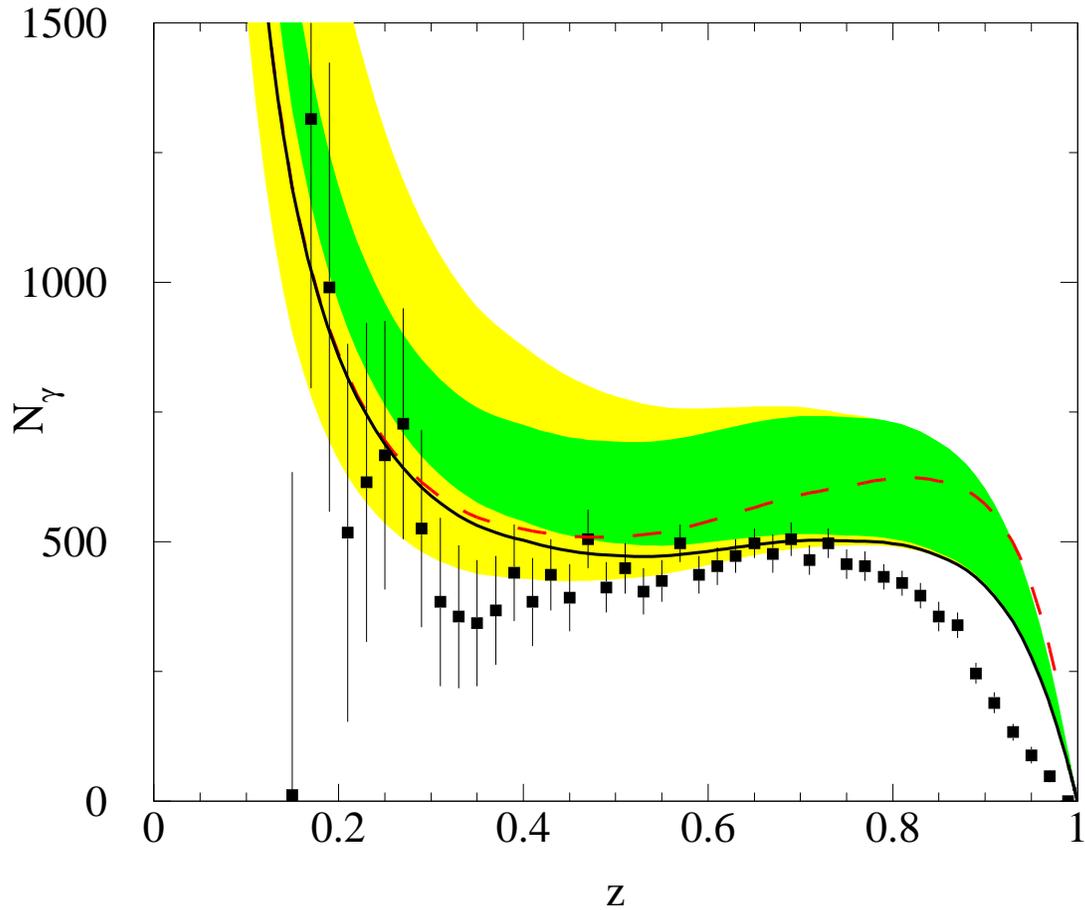}
\caption{\it The inclusive photon spectrum, compared with data
\cite{Nemati:1996xy}.  The theory predictions are described in the
text.}
\label{comparedata}
\end{figure}
in Fig.~\ref{comparedata}, along with the data.  The
error bars on the data are statistical only.  The dashed line is the
direct NRQCD tree-level plus fragmentation result, while the solid curve
includes NRQCD and SCET. For these two
curves the $\alpha_s$ extracted from these data was used:
$\alpha_s(M_\Upsilon) = 0.163$, which corresponds to $\alpha_s(M_Z) =
0.110$ \cite{Nemati:1996xy}.  The shape of the resummed result is much
closer to the data than the NRQCD only curve, though it is not a perfect fit. Also shown is the NRQCD + SCET result using the PDG value of $\alpha_s(M_Z)$~\cite{Hagiwara:pw}, including theoretical uncertainties, denoted by the shaded region.  The darker band is an estimate of theoretical errors, while the lighter band includes experimental errors related to the extraction of the photon fragmentation function~\cite{Buskulic:1995au}.


\begin{thebibliography}{0}

\bibitem{Weinberg:1978kz}
S.~Weinberg,
PhysicaA {\bf 96}, 327 (1979).

\bibitem{Witten:kx}
E.~Witten,
Nucl.\ Phys.\ B {\bf 122}, 109 (1977);

\bibitem{Georgi:qn}
H.~Georgi,
Ann.\ Rev.\ Nucl.\ Part.\ Sci.\  {\bf 43} (1993) 209;

\bibitem{Harvey:ya}
J.~A.~Harvey and J.~Polchinski,
``Recent Directions In Particle Theory: From Superstrings And Black Holes To The Standard Model. Proceedings, Theoretical Advanced Study Institute In Elementary Particle Physics, Boulder, Usa, June 1-26, 1992,''

\bibitem{Manohar:1995xr}
A.~V.~Manohar,
arXiv:hep-ph/9508245;

\bibitem{Kaplan:1995uv}
D.~B.~Kaplan,
arXiv:nucl-th/9506035.


\bibitem{Bauer:2000ew}
C.~W.~Bauer, S.~Fleming and M.~E.~Luke,
Phys.\ Rev.\ D {\bf 63}, 014006 (2001)
[arXiv:hep-ph/0005275].

\bibitem{Bauer:2000yr}
C.~W.~Bauer, S.~Fleming, D.~Pirjol and I.~W.~Stewart,
Phys.\ Rev.\ D {\bf 63}, 114020 (2001)
[arXiv:hep-ph/0011336].

\bibitem{Bauer:2001ct}
C.~W.~Bauer and I.~W.~Stewart,
Phys.\ Lett.\ B {\bf 516}, 134 (2001)
[arXiv:hep-ph/0107001].

\bibitem{Bauer:2001yt}
C.~W.~Bauer, D.~Pirjol and I.~W.~Stewart,
Phys.\ Rev.\ D {\bf 65}, 054022 (2002)
[arXiv:hep-ph/0109045].

\bibitem{Beneke:1}
M.~Beneke, A.~P.~Chapovsky, M.~Diehl and T.~Feldmann,
Nucl.\ Phys.\ B {\bf 643}, 431 (2002)
[arXiv:hep-ph/0206152]

\bibitem{Beneke:2}
M.~Beneke and T.~Feldmann,
Phys.\ Lett.\ B {\bf 553}, 267 (2003)
[arXiv:hep-ph/0211358]

\bibitem{Dugan:1990de}
M.~J.~Dugan and B.~Grinstein,
Phys.\ Lett.\ B {\bf 255}, 583 (1991).

\bibitem{Manohar:2002fd}
A.~V.~Manohar, T.~Mehen, D.~Pirjol and I.~W.~Stewart,
Phys.\ Lett.\ B {\bf 539}, 59 (2002)
[arXiv:hep-ph/0204229].

\bibitem{Chay:2002mw}
J.~g.~Chay and C.~Kim,
arXiv:hep-ph/0205117.


\bibitem{Mantry:2003uz}
  S.~Mantry, D.~Pirjol and I.~W.~Stewart,
  Phys.\ Rev.\ D {\bf 68}, 114009 (2003)
  [arXiv:hep-ph/0306254].

\bibitem{Mantry:2004pg}
  S.~Mantry,
  Phys.\ Rev.\ D {\bf 70}, 114006 (2004)
  [arXiv:hep-ph/0405290].

\bibitem{Blechman:2004vc}
  A.~E.~Blechman, S.~Mantry and I.~W.~Stewart,
  Phys.\ Lett.\ B {\bf 608}, 77 (2005)
  [arXiv:hep-ph/0410312].

\bibitem{Leibovich:2003tw}
  A.~K.~Leibovich, Z.~Ligeti, I.~W.~Stewart and M.~B.~Wise,
  Phys.\ Lett.\ B {\bf 586}, 337 (2004)
  [arXiv:hep-ph/0312319].

\bibitem{Coan:2001ei}
  T.~E.~Coan {\it et al.}  [CLEO Collaboration],
  Phys.\ Rev.\ Lett.\  {\bf 88}, 062001 (2002)
  [arXiv:hep-ex/0110055].

\bibitem{Abe:2001zi}
  K.~Abe {\it et al.}  [BELLE Collaboration],
  Phys.\ Rev.\ Lett.\  {\bf 88}, 052002 (2002)
  [arXiv:hep-ex/0109021].


\bibitem{Bauer:2001rh}
  C.~W.~Bauer, C.~W.~Chiang, S.~Fleming, A.~K.~Leibovich and I.~Low,
  Phys.\ Rev.\ D {\bf 64}, 114014 (2001)
  [arXiv:hep-ph/0106316].

\bibitem{Fleming:2002rv}
  S.~Fleming and A.~K.~Leibovich,
  Phys.\ Rev.\ Lett.\  {\bf 90}, 032001 (2003)
  [arXiv:hep-ph/0211303].

\bibitem{Fleming:2002sr}
  S.~Fleming and A.~K.~Leibovich,
  Phys.\ Rev.\ D {\bf 67}, 074035 (2003)
  [arXiv:hep-ph/0212094].

\bibitem{Fleming:2004rk}
  S.~Fleming and A.~K.~Leibovich,
  Phys.\ Rev.\ D {\bf 70}, 094016 (2004)
  [arXiv:hep-ph/0407259].

\bibitem{Fleming:2004hc}
  S.~Fleming, C.~Lee and A.~K.~Leibovich,
  Phys.\ Rev.\ D {\bf 71}, 074002 (2005)
  [arXiv:hep-ph/0411180].

\bibitem{firstRad}
S.~J.~Brodsky {\it et al.}, 
Phys.\ Lett.\ B {\bf 73}, 203 (1978);
K.~Koller and T.~Walsh,
Nucl.\ Phys.\ B {\bf 140}, 449 (1978).

\bibitem{bbl}
G.~T.~Bodwin {\it et al.},
Phys.\ Rev.\ D {\bf 51}, 1125 (1995)
[Erratum-ibid.\ D {\bf 55}, 5853 (1995)];
M.~E.~Luke {\it et al.},
Phys.\ Rev.\ D {\bf 61}, 074025 (2000).

\bibitem{Catani:1995iz}
S.~Catani and F.~Hautmann,
Nucl.\ Phys.\ Proc.\ Suppl.\  {\bf 39BC}, 359 (1995).

\bibitem{Maltoni:1999nh}
F.~Maltoni and A.~Petrelli,
Phys.\ Rev.\ D {\bf 59}, 074006 (1999).

\bibitem{Rothstein:1997ac}
I.~Z.~Rothstein and M.~B.~Wise,
Phys.\ Lett.\ B {\bf 402}, 346 (1997).

\bibitem{Kramer:1999bf}
M.~Kramer,
Phys.\ Rev.\ D {\bf 60}, 111503 (1999).

\bibitem{Nemati:1996xy}
B.~Nemati {\it et al.}  [CLEO Collaboration],
Phys.\ Rev.\ D {\bf 55}, 5273 (1997).

\bibitem{Hagiwara:pw}
K.~Hagiwara {\it et al.}  [Particle Data Group Collaboration],
Phys.\ Rev.\ D {\bf 66}, 010001 (2002).

\bibitem{Buskulic:1995au}
D.~Buskulic {\it et al.}  [ALEPH Collaboration],
Z.\ Phys.\ C {\bf 69}, 365 (1996).

\end{thebibliography}
\end{document}